# Pythagorean Triples and Cryptographic Coding

Subhash Kak
Oklahoma State University, Stillwater

**Abstract:** This paper summarizes basic properties of PPTs and shows that each PPT belongs to one of six different classes. Mapping an ordered sequence of PPTs into a corresponding sequence of these six classes makes it possible to use them in cryptography. We pose problems whose solution would facilitate such cryptographic application.

## Introduction

A Pythagorean triple ($a, b, c$) consists of positive integers that are the sides of a right triangle and thus $a^2 + b^2 = c^2$. Given a Pythagorean triple ($a, b, c$), we have other similar triples that are $d(a, b, c)$, where $d > 1$. A primitive Pythagorean triple (PPT) consists of numbers that are relatively prime.

Pythagorean triples have been found on cuneiform tablets of Babylon [1] and they are important in Vedic ritual and described in early geometry books of India [2]-[5] and in the works of Euclid and Diophantus.

For a PPT, $a,b,c$ cannot all be even. Also, $a, b$ cannot both be odd and $c$ even, because then $a^2 + b^2$ is divisible by 2, whereas $c^2$ is divisible by 4. One of $a$ and $b$ must, therefore, be odd, and we will use the convention that $b$ is even. Also note that the factors $(c - b)$ and $(c+b)$ of $(c^2 - b^2)$ must both be squares because they cannot have common factors other than 1 for otherwise they would not be primitive. If $c+b = s^2$ and $c-b=t^2$, where $s$ and $t$ are different odd integers with no common factors, solving them yields:

$$a = st,$$
$$b = \frac{s^2 - t^2}{2},$$
$$c = \frac{s^2 + t^2}{2}$$

This implies that there exists an infinity of PPTs. If the coordinate $(a/c, b/c)$ is seen as a point on the unit circle, it means that a countably infinity of these points are rational. Note further that every odd number is a difference of the squares of two consecutive numbers and a subset of them is perfect squares. A sequence that generates a subset of PPTs is $(2n+1, 2n^2+2n, 2n^2+2n+1)$ for $n = 1,2,3…$

PPTs may be indexed in a variety of ways. It was shown by Barning [6],[7] that they may be generated in a tree starting with the triple (3,4,5) using the three matrix transformations below:

$$T_1 = \begin{bmatrix} -1 & 2 & 2 \\ -2 & 1 & 2 \\ -2 & 2 & 3 \end{bmatrix}, T_2 = \begin{bmatrix} 1 & 2 & 2 \\ 2 & 1 & 2 \\ 2 & 2 & 3 \end{bmatrix}, T_3 = \begin{bmatrix} 1 & -2 & 2 \\ 2 & -1 & 2 \\ 2 & -2 & 3 \end{bmatrix}$$

If *a=st* in the present generation, in the next generation, the values of *a* for the three successors will be:

*s (2s – t), s (2s + t), t (s + 2t)*

For the triple *(3,4,5)* for which *s=3, t=1*, the 3 values of *a* in the next generation are 15, 21, and 5, respectively. The Barning tree for the first three generations is given in Figure 1 below:

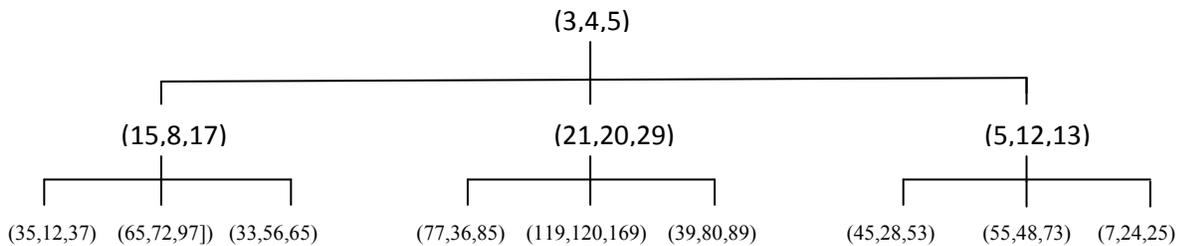

**Figure 1.** First three generations of primitive Pythagorean triples

In this paper we summarize basic properties of PPTs and show that each PPT belongs to one of six different classes. It is shown how an ordered sequence of PPTs transformed into these six classes could have use in cryptography. We pose problems whose solution would facilitate their cryptographic application.

## Brief History

The earliest statement of the theorem of the square on the diagonal (Pythagoras theorem), together with some examples, is to be found in the geometry text of Baudhāyana (c. 800 BC). In his Śulba Sūtra 1.12 and 1.13, it is stated [2]:

> The areas (of the squares) produced separately by the length and the breadth of a rectangle together equal the area (of the square) produced by the diagonal. This is observed in rectangles having sides 3 and 4, 12 and 5, 15 and 8, 7 and 24, 12 and 35, and 15 and 36.

O'Conner and Robertson in their history of mathematics project take Baudhāyana to be 800 BC [8]. Seidenberg [9] presents various arguments for an early date for the knowledge of the Pythagoras theorem in India (for other aspects of early Indian mathematics, see [10]-[12] and for the altar ritual that provides the context in which this mathematics was used in the Śulba Sūtras, see [13],[14]). Van der Waerden saw a ritual origin to the discovery of Pythagoras [15].

The Śulba Sūtras are texts of applied geometry that provide techniques to draw altars of different shapes and sizes in a convenient manner. The word *śulba* means a "cord", "rope", or "string" and the root *śulb* signifies "measurement". The cord has marks (*nyañcana* in Sanskrit) that indicate where the intermediate pegs are to be fixed. Thus a cord of 12 units length with *nyañcana* at 3 and 7, can be readily stretched to yield the right-angled triangle (3,4,5).

It is significant that of the six examples given by Baudhāyana, five are primitive triples:

   3,   4,   5

  12,   5,  13

  15,   8,  17

   7, 24, 25

  12, 35, 37

The only non-primitive triple in the list, namely (15,36, 39) seems to have been included because it was widely known to the readers of the Śulba Sūtra as it is fundamental to the design of the Mahāvedi altar of the Vedic ritual. The significance of the (15,36,39) triple derives from the fact that the sum of the three numbers is 90 (one-fourth the days in the year) which equals the size of the altar (base = 30, height =36, and top =24) (Figure 2). Although the construction of the Mahāvedi altar in the older Śatapatha Brāhmaṇa does not directly mention the use of *nyañcana* of the cord, it is clear how the (15,36,39) triple would serve as an excellent check on the altar dimensions traced on the ground.

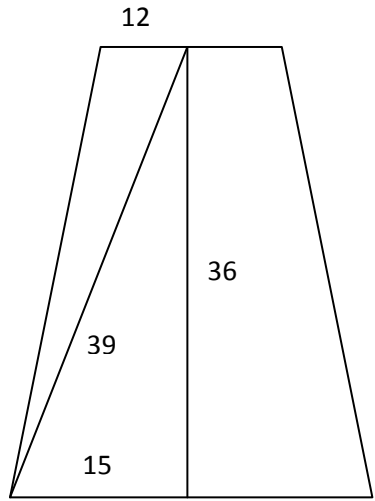

**Figure 2.** The Mahāvedi altar (base 30, height 36, top 24)

In the altar of Figure 2, the triple (12,35,37) could also be used as a check on the accuracy of the construction.

## Representing Pythagorean Triples as Gopal-Hemachandra Numbers

Consider the Gopala-Hemachandra (GH) quadruple *(g,e,f,h)*. The GH sequence, named after two mathematicians who lived before Fibonacci [16], is the sequence

    *g, e, g+e, g+2e, 2g+3e, 3g+5e, ...*

for any pair *g, e*. When *g=1, e=1,* we obtain the Fibonacci sequence.

**Theorem 1.** *In the GH quadruple (g,e,f,h), if* $a = gh$, $b = \dfrac{(c-a)f}{e}$, *and* $c = c = eh + fg$, *then (a, b, c) is a Pythagorean triple. If the quadruple has no common factors and g is odd, then (a,b,c) is a PPT. The values b, and c may also be written as b=2ef, and c= $e^2+f^2$.*

*Proof.* By definition of GH numbers, *f=g+e* and *h= e+f = g+2e*. By substitution, $c=2e^2+g^2+2ge$, $a=g^2+2ge$, $b=2ge+2e^2$, and $a^2 + b^2 = c^2$, ensuring that the quadruple represents a PT. When *g* is even, *h* is also even, which implies that *c* is even and the triple is not primitive. The alternate values of *b* and *c* are obtained by substitution.

This means that *a* is obtained by multiplying the outer two numbers in the *(g,e,f,h)* representation, *b* is twice the product of the inner two numbers, and *c* is the sum of the squares of the inner two numbers. For the quadruple (1,1,2,3), $c=3+2=1^2+2^2=5$, $a=3\times1=3$, and $b=2\times2=4$.

**Corollary.** *The GH quadruple may be written in the form* $(t, \dfrac{s-t}{2}, \dfrac{s+t}{2}, s)$ *where s and t are distinct odd integers, and s > t.*

**Algorithm for generating Barning tree using GH numbers**. Using the GH quadruple *(g,e,f,h)*, the next Barning generation may be generated by the algorithm:

    Left child: *(h,e,h+e,h+2e)*

    Middle child: *(h,f,h+f,h+2f)*

    Right child: *(g,g+e,2g+e,3g+2e)*

Notice that the smallest GH quadruple is (1,1,2,3). The next smallest GH quadruples are obtained from the numbers 3,1; 3,2; and 1,2 in one arrangement as shown below:

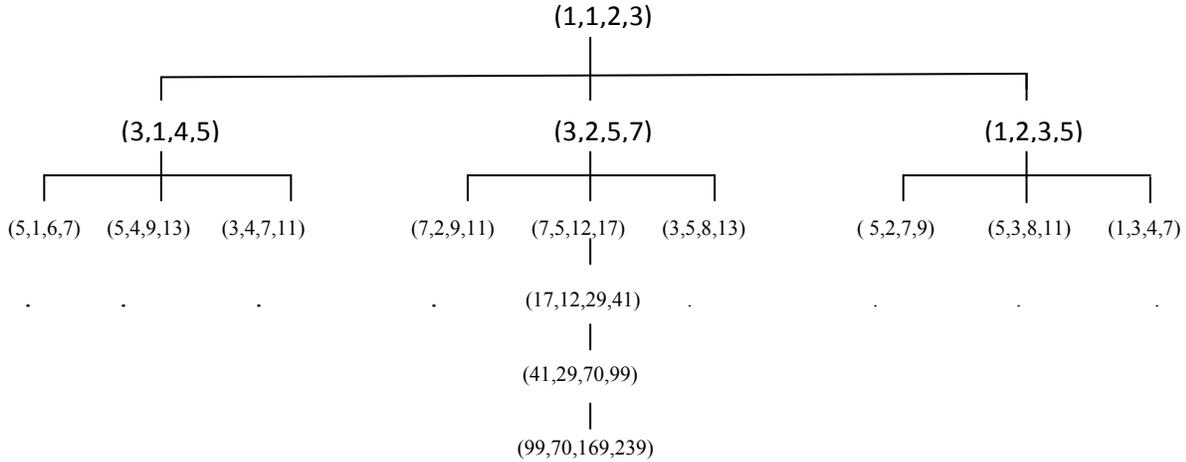

**Figure 3.** Six generations showing the middle child of the PPT tree

The quadruple associated with 1,3 shows up in the next generation and, therefore, we have accounted for all the immediate neighbors of (1,1,2,3). The GH numbers corresponding to the PPTs in Generation 2 are spanned by (3,1,4,5) and (1,2,3,5); in Generation 3 by (5,1,6,7) and (1,3,4,7); in Generation 4 by (7,1,8,9) and (1,4,5,9), and in Generation *n* by *(2n-1,1,2n,2n+1)* and *(1,n,n+1,2n+1)*.

Looking at Figure 3, if we consider the middle child in sequence across generations using quadruples, we have the sequence:

$$\begin{array}{cccc} a & b & c & d \\ d & c & d+c & d+2c \\ d+2c & d+c & 2d+3c & 3d+4c \\ 3d+4c & 2d+3c & 5d+7c & 7d+10c \\ 7d+10c & 5d+7c & 12d+17c & 17d+24c \\ \dots \end{array}$$

In the last column, if we call the *n*th term as *d(n)*, then *d(n)=2d(n-1)+d(n-2)*. The ratio $r = \lim_{n \to \infty} \frac{d_n}{d_{n-1}}$ is obtained by dividing both sides by *d(n-1)* and using the limit which gives us this value as solution to the equation *r² - 2r -1 =0*. Or,

$$r = \frac{2+\sqrt{5}}{2} \cong 2.118$$

The related golden ratio (Φ = 1.618..), solution to the algebraic equation *x² – x – 1 = 0,* is the limiting ratio of the Fibonacci series 1, 1, 2, 3, 5, … . The Fibonacci series is itself derived easily from the Mount Meru (*Meru-Prastāra*) of Piṅgala (c. 200 BC but perhaps 450 BC if the many textual notices about him being the younger brother of the great grammarian Pāṇini are right) as shown in Figure 4.

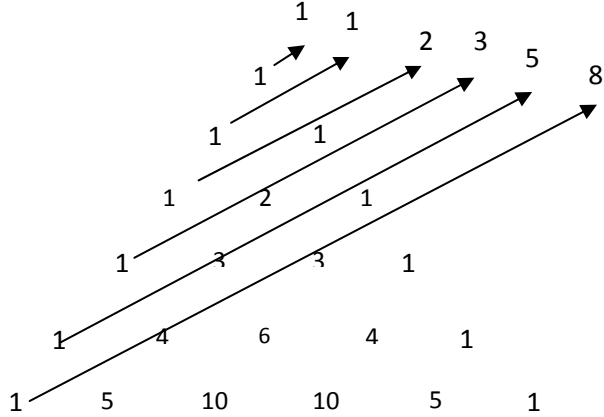

**Figure 4.** Fibonacci sequence obtained from the Meru Prastāra

There exists a variant to the Barning tree for generating PPTs [17] that will not be discussed here.

## Indexing Using s and t Numbers

A more convenient indexing than obtained using the Barning tree is where relatively prime *s* and *t* numbers are used in an array where $s > t$. This may be seen in the diagram shown below:

|   |    | s |   |   |   |   |   |   |   |   |
|---|----|---|---|---|---|---|---|---|---|---|
|   |    | 3 | 5 | 7 | 9 | 11 | 13 | . | . | . |
|   | 1  | (1,3) | (1,5) | (1,7) | (1,9) | (1,11) | (1,13) | . | . | . |
|   | 3  |       | (3,5) | (3,7) | .     | (3,11) | (3,13) | . | . | . |
|   | 5  |       |       | (5,7) | (5,9) | (5,11) | (5,13) | . | . | . |
| t | 7  |       |       |       | (7,9) | (7,11) | (7,13) | . | . | . |
|   | 9  |       |       |       |       | (9,11) | (9,13) | . | . | . |
|   | 11 |       |       |       |       |        | (11,13)| . | . | . |
|   | .  |       |       |       |       |        |        | . | . | . |
|   | .  |       |       |       |       |        |        | . | . | . |

**Table 1.** Array of PPTs ordered by *s* and *t* numbers

If $s \leq n$, and $u=(n-1)/2$, then the total number of PPTs, $\vartheta(n)$, will be bounded by $u(u+1)/2$. The PPTs may now be indexed in a variety of ways. One may go across rows or columns or across diagonals of different slope.

Some values of $\vartheta(n)$ are given below:
$\vartheta(3) = 1, \vartheta(5) = 3, \vartheta(7) = 6, \vartheta(9) = 9$
$\vartheta(11) = 14, \vartheta(13) = 20, \vartheta(15) = 24, \vartheta(17)=32,...$

If the indexing were done according to columns of Table 1, we will have the following PPTs in the array across the first seven generations:

(3,4,5)
(5,12,13)   (15,8,17)
(7,24,25)   (21,20,29)   (35,12,37)
(9,40,41)   (45,28,53)   (63,16,65)
(11,60,61)  (33,56,65)   (55,48,73)   (77,36,85)   (99,20,101)
(13,84,85)  (39,80,89)   (65,72,97)   (91,60,109)  (117,44,125)  (143,24,145)
(15,112,113)            (105,88,137) (165,52,173) (195,28,197)
.   .   .   .   .   .   .
.   .   .   .   .   .   .

**Table 2.** Example generations of PPTs ordered by *s* and *t* numbers

## Cryptographic Coding

**Theorem 2.** *The Pythagorean triples satisfy the property that the numbers 3, 4, and 5 divide a, b, c either separately or jointly in different ways.*

*Proof.* As before $a = st$, $b = (s^2 - t^2)/2$, and $c = (s^2 + t^2)/2$. If 3 divides *s* or *t*, then it also divides *a*. If $s = \pm 1$ mod 3 and $t = \pm 1$ mod 3, then $s^2 - t^2 = 0$ mod 3, so that is *b* is divisible by 3. Since both *s* and *t* are odd and different, they are both $\pm 1$ mod 4, implying $b = 0$ mod 4. When neither *s* nor *t* is divisible by 5, they can only be $\pm 1$ or $\pm 2$ mod 5, which means that either $s^2 - t^2$ or $s^2 + t^2$ is 0 mod 5. Considering all these cases for *c*, it is clear that it is never divisible by 3.

**Theorem 3.** *Primitive Pythagorean triples come in 6 classes based on the divisibility of a, b, c by 3, 4, and 5.*

*Proof.* We can ignore divisibility by 4, for that always characterizes *b*. Now assume 3 divides *a*, then 5 can divide either *a, b,* or *c*. Next assume that 3 divides *b*, then 5 can divide *a*, *b*, or *c*. This enumerates all the six possibilities. The 6 classes are listed out as follows.

1. Class A, in which *a* is divisible by 3 and *c* is divisible by 5.

| div | a | b | c |
|---|---|---|---|
| 3 | X | | |
| 5 | | | X |

Examples: (3,4,5), (33,56,65)

2. Class B, in which *a* is divisible by 5, and *b* is divisible by 3.

   | div | a | b | c |
   |-----|---|---|---|
   | 3   |   | X |   |
   | 5   | X |   |   |

   Examples: (5,12,13), (35,12,37)

3. Class C, in which *a* is divisible by 3 and 5.

   | div | a | b | c |
   |-----|---|---|---|
   | 3   | X |   |   |
   | 5   | X |   |   |

   Examples: (15,8,17), (45,28,53)

4. Class D, in which *b* is divisible by 3 and *c* is divisible by 5.

   | div | a | b | c |
   |-----|---|---|---|
   | 3   |   | X |   |
   | 5   |   |   | X |

   Examples: (7,24,25), (13,84,85)

5. Class E, in which *a* is divisible by 3 and *b* is divisible by 5.

   | div | a | b | c |
   |-----|---|---|---|
   | 3   | X |   |   |
   | 5   |   | X |   |

   Examples: (21,20,29), (9,40,41)

6. Class F, in which b is divisible by 3 and 5.

   | div | a | b | c |
   |-----|---|---|---|
   | 3   |   | X |   |
   | 5   |   | X |   |

   Examples: (11,60,61), (91,60,109)

The six different classes, with the labels A,B,C,D,E,F, in combinations of two define 36 possibilities that will suffice to represent the English alphabet (without case distinction) together with a few items for space and certain punctuations.

Let the mapping of a PPT, *x*, into the corresponding class be represented by the function *w(x)*. The mapping w(x) for the first 33 PPTs (ordered in terms of the magnitude of the largest term) is:

| PPT | | | w(x) |
|---:|---:|---:|:---:|
| 3 | 4 | 5 | A |
| 5 | 12 | 13 | B |
| 15 | 8 | 17 | C |
| 7 | 24 | 25 | D |
| 21 | 20 | 29 | E |
| 35 | 12 | 37 | B |
| 9 | 40 | 41 | E |
| 45 | 28 | 53 | C |
| 11 | 60 | 61 | F |
| 33 | 56 | 65 | A |
| 63 | 16 | 65 | A |
| 55 | 48 | 73 | B |
| 13 | 84 | 85 | D |
| 77 | 36 | 85 | D |
| 39 | 80 | 89 | E |
| 65 | 72 | 97 | B |
| 99 | 20 | 101 | E |
| 91 | 60 | 109 | F |
| 15 | 112 | 113 | C |
| 117 | 44 | 125 | A |
| 105 | 88 | 137 | C |
| 17 | 144 | 145 | D |
| 143 | 24 | 145 | D |
| 51 | 140 | 149 | E |
| 85 | 132 | 157 | B |
| 119 | 120 | 169 | F |
| 165 | 52 | 173 | C |
| 19 | 180 | 181 | F |
| 57 | 176 | 185 | A |
| 153 | 104 | 185 | A |
| 95 | 168 | 193 | B |
| 195 | 28 | 197 | C |
| 133 | 156 | 205 | D |

If the classes A,B,C,D,E,F are mapped into the digits 0,1,2,3,4,5, the sequence of PPTs ordered by increasing largest term may be written as the irrational number

$$0.012341425001334145202334152500123\ldots.$$

If there are several PPTs for the same largest term, as in the example of (33,56,65) and (63,16,65), they are further ordered by the increasing *a* term.

## Two Problems for Cryptographic Application of w subsequences

Given a choice of a PPT and its immediate successors that we call $x_n, x_{n+1}, \ldots$, one may write down the corresponding $w(x)$s. For example, in the list of 33 PPTs given above, the four-unit long subsequences are: ABCD, BCDE, CDEB, DEBE, EBEC, and so on. One may easily check that for the first 13 PPTs, a subsequence of length 4 suffices for unique identification of each of the PPTs.

Starting with an indexing scheme for PPTs (Barning tree or indexing using *s* and *t* numbers, or some other scheme), one can generate w-subsequences of suitable length to map messages. But for this to become an effective cryptographic scheme, it is essential to solve the following problems:

1. What is the minimum size, *i,* of a subsequence $w(x_n), w(x_{n+1}), \ldots, w(x_{n+i-1})$ so that $x_n$ less than a certain value (in terms of the number of elements in the indexed set) can be uniquely identified? A related problem is: how does the value of *i* vary with *n*?
2. Is there an efficient way to compute $x_n$ given $w(x_n), w(x_{n+1}), \ldots, w(x_{n+i-1})$ ?

The solution to these problems would help determine the cryptographic complexity of the w-sequences.

## Conclusions

This paper is an introduction to the use of PPTs in cryptography. General properties of PPTs are summarized; their description in Indian texts and the context in which they were used is given. It is shown that each PPT belongs to one of six distinct classes and it is proposed that this property be used for cryptographic applications. Two problems are posed whose solution would facilitate the application of PPTs to cryptography.

Stillwater
April 21, 2010